\documentclass[12pt, A4]{article}
\usepackage[margin=1in]{geometry}  
\usepackage[utf8]{inputenc}  

\usepackage{graphicx}
\graphicspath {{Figures/}}

\usepackage{epstopdf} 
\usepackage{placeins} 
\usepackage{caption}
\usepackage{subcaption}  
\usepackage{multirow} 
\usepackage{array}  
\usepackage{float} 
\usepackage{bm}
\usepackage{amssymb} 
\usepackage{mathtools}
\usepackage{authblk}
\usepackage{color}
\usepackage{setspace}   
\usepackage{comment}
\usepackage{soul} 
\usepackage{vector}  
\usepackage[backend=bibtex, style=numeric-comp, sorting=none, maxbibnames=99]{biblatex}
\title{A brief introduction to bulk viscosity of fluids}
\author{Bhanuday Sharma, Rakesh Kumar}

\date{%
	Department of Aerospace Engineering, Indian Institute of Technology Kanpur, India\\[2ex]%
	\today
}

\begin{document}
\maketitle
\setstretch{1.2}  
\begin{abstract}
	Fluid flows are typically studied by solving the Navier--Stokes equation. One of the fundamental assumptions of this equation is Stokes’ hypothesis. This hypothesis assumes bulk viscosity, $\mu_b$, to be identically zero. The Stokes' hypothesis is a reasonable approximation for commonly observed fluid flows; therefore, Navier--Stokes equation gives satisfactory results in these situations. However, there are circumstances where this hypothesis does not hold good, and therefore, the classical Navier--Stokes equation becomes inapt. These situations include the absorption of sound waves, hypersonic flows, turbulent flows, and flow of Martian air etc. Reliable analytical and computational studies of these flows requires account of bulk viscosity in the governing equations. In this article, we provide a brief review of the subject of bulk viscosity. We start with a brief background of this topic. Then we discuss the underlying microscopic mechanisms that give rise to bulk viscosity effects. It was followed by a review of methods available in the literature for estimation of this parameter. Finally, a review of the studies that analyze the effects of bulk viscosity in various fluid flows is provided.
	
\end{abstract}




\section{Introduction}

Ever since Sir George Gabriel Stokes (1819-1903) proposed the complete set of equations for the dynamics of viscous fluids in 1845 \cite{stokes1880theories}, bulk viscosity (sometimes also referred as volume or  dilatational viscosity) has remained as one of the controversial subjects of fluid dynamics \cite{gad1995questions}. The recent surge of interest in Mars missions has made the study of bulk viscosity more relevant as the Martian atmosphere consists approximately 96\% of carbon dioxide gas, which has a reported bulk to shear viscosity ratio of $\approx 2000$~\cite{tisza1942supersonic, truesdell1954present}. The account of bulk viscosity in the analysis has also enabled more accurate modeling of several fluid mechanical phenomena \cite{emanuel1992effect, gonzalez1993effect, emanuel1994linear, orou1994second, xu2006continuum, elizarova2007numerical, fru2012impact, cramer2014effect, bahmani2014suppression, chikitkin2015effect, sengupta2016roles, singh2017computational, lin2017bulk}. On the other hand, in contrast to shear viscosity, which is a very well studied transport property, bulk viscosity is still not completely explored. There are considerable ambiguities/uncertainties about the nature, effects, and applicability of the concept of bulk viscosity. Even for most common fluids, existing experimental values of bulk viscosity are spread over a broad range, and widely accepted values are still not available\cite{marcy1990evaluating}. 
In this article, we provide a brief review of the subject of bulk viscosity. We start with a brief background of this subject. Then we discuss the underlying microscopic mechanisms that give rise to bulk viscosity effects. It was followed by a review of methods available in the literature for estimation of this parameter. Finally, a review of the studies that analyze the effects of bulk viscosity in various fluid flows is provided.

The stress-strain rate relationship (i.e., constitutive relation) for a Newtonian fluid is given as follows:
\begin{equation}
\sigma_{ik}=-P_{thermo}~\delta_{ik}+\mu\displaystyle\left(\frac{\partial u_i}{\partial x_k}+\frac{\partial u_k}{\partial x_i}\right)+\lambda \frac{\partial u_j}{\partial x_j} \delta_{ik}
\label{Constitutive_Relation_in_lambda}
\end{equation}
where, $\sigma_{ik}$, is Cauchy's stress tensor, $\delta_{ik}$ is the Kronecker delta, $u_i$ is velocity of the fluid, $x_i$ is spatial coordinate, and the scalar quantity $P_{thermo}$ is the thermodynamic pressure or hydrostatic pressure. The above relation contains two independent coefficients - first is the coefficient of shear viscosity ($\mu$), which is sometimes also termed as the first coefficient of viscosity, and the second is the coefficient of longitudinal viscosity ($\lambda$), which is also referred to as the second coefficient of viscosity.

The above relation, Eq.~(\ref{Constitutive_Relation_in_lambda}), can be rearranged as follows by separating isotropic and deviatoric parts of the strain-rate tensor:
\begin{equation}
\sigma_{ik}=-P_{thermo}~\delta_{ik}+\mu\displaystyle\left(\frac{\partial u_i}{\partial x_k}+\frac{\partial u_k}{\partial x_i} - \frac{2}{3} \frac{\partial u_j}{\partial x_j} \delta_{ik} \right)+\mu_b \frac{\partial u_j}{\partial x_j} \delta_{ik}
\label{constitutive_relation}
\end{equation}
where, $\mu_b=\displaystyle\left(\frac{2}{3}\mu + \lambda \right)$. The coefficient $\mu_b$ is known as the {coefficient of bulk viscosity}, and it represents the irreversible resistance, over and above the reversible resistance, caused by isentropic bulk modulus to change of volume \cite{hoover1980bulk}. Its values are expressed in the same units as shear viscosity, i.e., Pa~s or poise. The second law of thermodynamics constrains both the shear viscosity and bulk viscosity to have only non-negative values~\cite{landau1959lifshitz}. The values of bulk viscosity for common dilute gases at 300 K are listed in Table~\ref{tab:mu_b from crammer}~\cite{cramer2012numerical}. It should be noted that the bulk viscosity of monatomic gases in the dilute gas limit is zero.

\begin{table}[h!]
  \begin{center}
        \begin{tabular}{|c|c|c|c|}
            \hline \rule[-2ex]{0pt}{5.5ex} \textbf{Gas} & \textbf{$\bold{\mu_b/\mu}$} & \textbf{Gas}  &  \textbf{$\bold{\mu_b/\mu}$}  \\ 
            \hline \hline \rule[-2ex]{0pt}{5.5ex}  Carbon monoxide & 0.548  & Dimethylpropane  &  3.265 \\ 
            \hline \rule[-2ex]{0pt}{5.5ex}  Nitrogen & 0.769  & Water vapor  & 7.36 \\ 
            \hline \rule[-2ex]{0pt}{5.5ex}  n-Pentane & 0.896  & Hydrogen & 28.95  \\
            \hline \rule[-2ex]{0pt}{5.5ex}  iso-Pentane & 1.057  & Chlorine & 751.88  \\ 
            \hline \rule[-2ex]{0pt}{5.5ex}  n-Butane & 1.13  & Fluorine & 2329  \\ 
            \hline \rule[-2ex]{0pt}{5.5ex}  iso-Butane & 2.00  & Carbon dioxide & 3828  \\  
            \hline 
        \end{tabular} 
        
      \caption{Ratio of bulk viscosity to shear viscosity for common gases at 300 K \cite{cramer2012numerical}}
      \label{tab:mu_b from crammer}
    \end{center}
\end{table}

The mechanical pressure, ${P}_{mech}$, is defined as the negative average of the diagonal terms of stress tensor, as given below:
\begin{equation}
{P}_{mech} = - \frac{1}{3}(\sigma_{11}+\sigma_{22}+\sigma_{33})= P_{thermo}-\mu_b \, \nabla \cdot\vec{u}
\label{Pmech=Pthermo+div}
\end{equation} 
Historically, Stokes~\cite{stokes1880theories} assumed the bulk viscosity ($\mu_b$) to be identically zero for all fluids. It implies that mechanical pressure is always equal to thermodynamic pressure, irrespective of the process through which the system is undergoing, i.e., the viscous forces do not depend upon the rate of expansion or compression at all. This assumption is known as the Stokes' hypothesis. Later, it became customary to use this hypothesis in fluid mechanics. However, Stokes~\cite{stokes1880theories} himself did not take this hypothesis as always true. He mentioned that in commonly encountered flows, if analysis with and without considering bulk viscosity produces the same results, it would be because of small $\nabla\cdot\vec{u}$ rather than $\mu_b$ being zero. Indeed, there are certain instances, where bulk viscosity effects are not negligible. The Section~\ref{Sec:Applications} briefly discusses these scenarios, and some of the past works to model them while considering finite bulk viscosity.

\section{Microscopic picture of mechanisms of generation of normal stresses in an expanding/contracting dilute gas} \label{Sec: Normal_stresses}

To understand the microscopic origin of the normal stress, let's re-write Eq.~\ref{constitutive_relation} for the normal stresses acting in the x-direction on yz-plane of the fluid element.
\begin{equation}
\sigma_{11}=-P_{thermo}+2\mu \displaystyle\left( \frac{\partial u_1}{\partial x_1}- \frac{1}{3} \nabla \cdot \vec{u}\right)+\mu_b \nabla \cdot \vec{u}
\end{equation}
The second and third term in the above equation are contributions of shear and bulk viscosity to normal stress, respectively. Mechanisms responsible for stresses due to these terms are discussed as follows.

\begin{figure}[h]
	\centering
	\includegraphics[width=0.25\columnwidth]{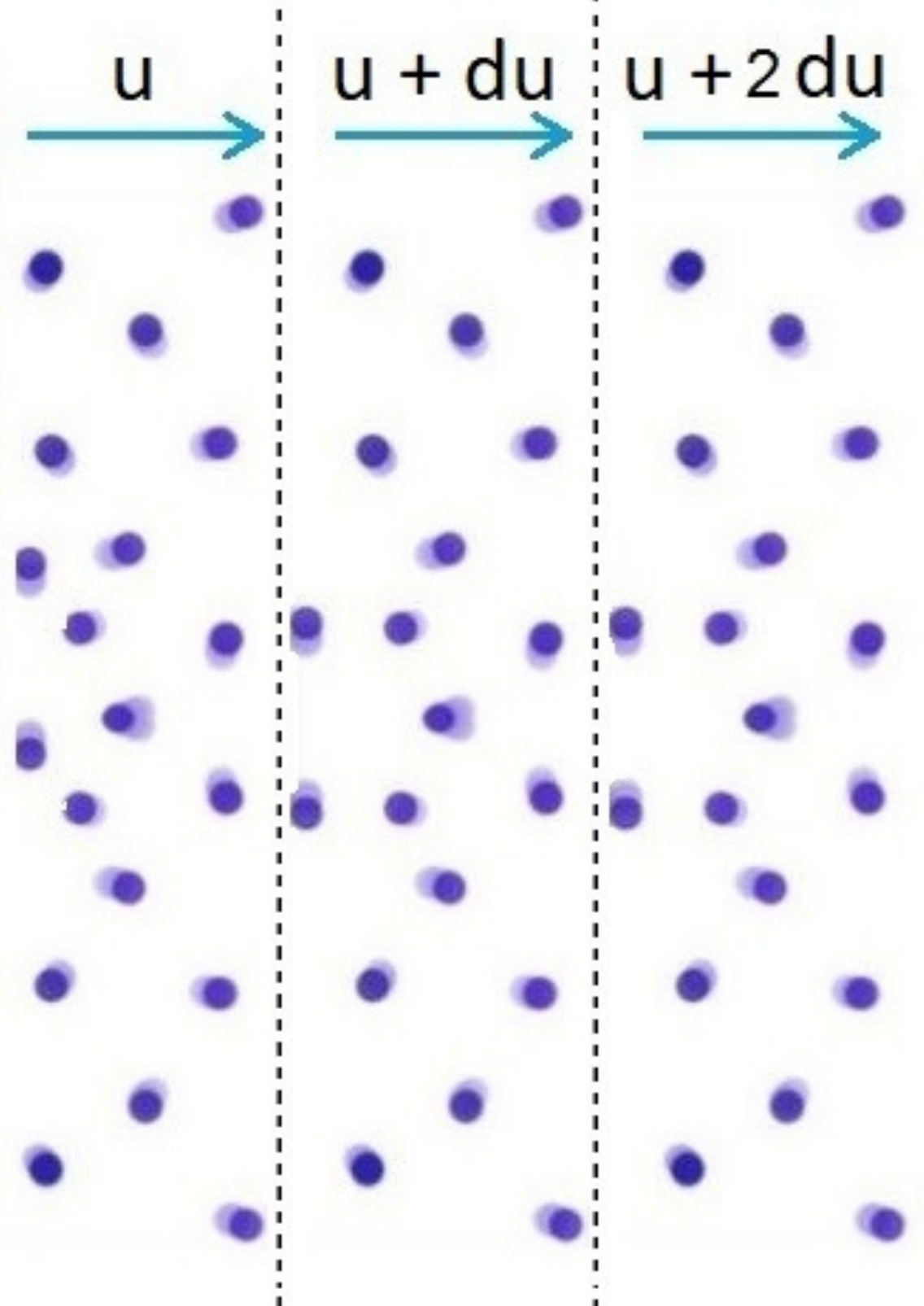}
	\caption{Three layers of different velocity in a divergent flow field. Dots represent gas molecules and two vertical dashed lines are imaginary boundaries separating these three fluid layers.}
	\label{3-layers}
\end{figure}

\subsection{Shear viscosity}

In dilute gases, the mechanism of generation of normal stress due to shear viscosity is analogous to the mechanism that produces shear stress, i.e.,  the transport of momentum in adjacent layers of fluid due to thermal diffusion of molecules.
Consider a flow field dilating only in one direction with three different layers of fluid having velocities, as shown in Fig.~\ref{3-layers}. The left, middle and right layers have a bulk velocity equal to $u$, $u+du$, and $u+2du$, respectively, in the direction as shown in Fig.~\ref{3-layers}. When any molecule from the leftmost layer jumps to middle layer due to its random-thermal-motion, it will decrease the average momentum of middle layer. As a result, middle layer will be pulled towards the left side. 	
Similarly, if any molecule from the rightmost layer jumps to the middle layer, it will increase the average momentum of middle layer. As a consequence, there acts a pull on the middle layer towards the right side.
The combined effect of these two pulls will be a normal stress on the fluid element of middle layer along the direction of velocity gradient. Therefore, in contrast to what the name `shear viscosity' can mislead to, shear viscosity can not only produce shear stresses but also normal stresses.

\subsection{Bulk viscosity} 
The pressure in any fluid is summation of two components--
\begin{equation}
	P = P_{\text{kinetic}} + P_{\text{virial}}
\end{equation}
The first contribution, i.e., $P_{\text{kinetic}}$ is represents force caused by bombardment of molecules on a unit area. The second contribution, $P_{\text{virial}}$ is caused by inter-molecular forces. It depends upon the strength of inter-molecular interaction. It should be noted that the $P$ can refer to both mechanical and thermodynamic pressure depending on the reference. The mechanical pressure, $P_{mech}$, is instanteous pressure of the fluid at all situations, i.e., both at equilibrium or non-equilibrium. Hydrostatic/thermodynamic pressure, $P_{thermo}$, is the mechanical pressure of the system under hydrostatic/equilibrium conditions, i.e., when fluid is at rest. Therefore, for a system in a state of non-equilibrium, the term `thermodynamic pressure' loses its meaning \cite{landau1959lifshitz}. However, for such a system, it can still be defined as the mechanical pressure of the system if the system is brought to equilibrium state adiabatically \cite{okumura2002new}. 

The origin of bulk viscosity in fluids is due to the fact that the pressure at non-equilibrium (i.e., mechanical pressure) is not same as the pressure at equilibrium (i.e., thermodynamic pressure). There are primarily two mechanisms responsible for deviation of equilibrium pressure from non-equilibrium pressure-~(a)~The finite rate of relaxation of internal degrees of freedom with random translational energy (b)~The finite rate of structural rearrangement of molecules after a change in the thermodynamic state of the fluid. The bulk viscosity due to first mechanism is called as `apparent bulk viscosity', while bulk viscosity due to the second mechanism is called as `intrinsic bulk viscosity' \cite{nettleton1958intrinsic}. The total bulk viscosity of fluid is summation due to the two said contributions.

\subsubsection{Apparent bulk viscosity}

\begin{figure}[htb]
    \centering
    \includegraphics[width=0.25\columnwidth]{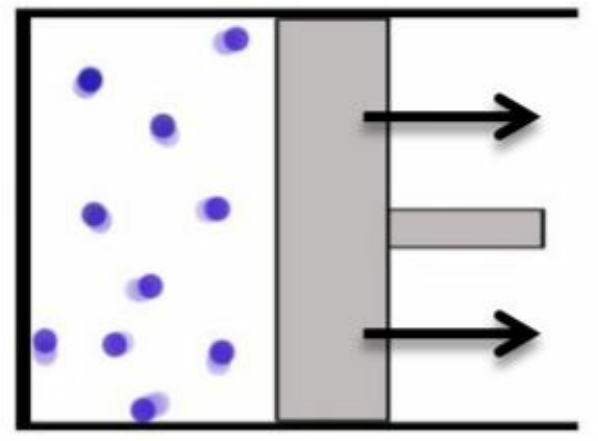}
    \caption{Expansion of gas in a piston-cylinder arrangement}
    \label{piston}
\end{figure}

The apparent bulk viscosity appears when the $P_{\text{kinetic}}$ at non-equilibrium is not same as $P_{\text{kinetic}}$ at equilibrium due to presence of internal degree of freedoms, i.e., (viz., rotational and vibrational).
Herzfeld and Rice \cite{herzfeld1928dispersion} first suggested that microscopic cause of bulk viscosity is the finite rate of exchange of energy between the translational mode and internal degrees of freedom. The mechanism can be explained by considering a simple example in which a polyatomic dilute gas, say nitrogen, is expanding adiabatically in a piston-cylinder arrangement, as shown in Fig.~\ref{piston}. Since the gas is considered as dilute, the virial pressure is neglected. 

Let us assume that initially the piston was at rest, and a polyatomic dilute gas was in equilibrium at a temperature of 300 K. With the assumption of inactive vibrational modes at the prevailing temperature conditions, the gas has three translational and two rotational degrees of freedom. Since the gas is in equilibrium, these all five degrees of freedom possess equal amount of energy because of the equipartition law of energy. Now, when the gas expands, it does work against the piston and loses its energy. The energy that the gas loses comes directly from its translational mode, whereas the energy associated with rotational mode remains momentarily unaffected. It causes an imbalance in the equipartition of energy among translational and rotational degrees of freedom. At this stage, the system is in a state of non-equilibrium, and its current translational kinetic energy is less than that which would be if the system is again brought to equilibrium adiabatically. Similarly, its rotational kinetic energy is more than that which would be if the system is brought to equilibrium. The system tries to regain its state such that the whole kinetic energy is equally distributed among all degrees of freedom. In this attempt, the system transfers some of its kinetic energy from internal modes to translational modes by means of inter-molecular collisions amongst gas molecules.

The mechanical pressure, $P_{mech}$, at any point in the fluid is the negative average of normal stresses acting on that point. For a dilute gas with negligible virial pressure, it represents actual force caused by bombardment of molecules on a unit area, and therefore, depends only upon the random translational kinetic energy of the gas molecules. This is also the reason why PV work comes at the cost of translational kinetic energy. Kinetic theory of gases relates mechanical pressure to translational kinetic energy ($E_{trans}$) by the following relation:
\begin{equation}
P_{{mech}} = \frac{2~E_{trans}}{3\mathcal{V}} \label{eq: P_mech}
\end{equation}
where $\mathcal{V}$ is volume of the system. 
Moreover, due to the equipartition law of energy, the translational kinetic energy ($E_{trans}$) of the system at equilibrium is equal to $3/f$ of the total energy ($E_{total}$), where $f$ is number of total degrees of freedom of gas. Thus, the thermodynamic pressure can be given as follows:
\begin{eqnarray}
P_{thermo }&=& \frac {2}{3 \mathcal{V}} \displaystyle \left(E_{trans}~\text{at equilibrium} \right) = \frac {2}{3 \mathcal{V}}  \left(\frac{3}{f} E_{total} \right) \\
P_{{thermo}} &=& \frac{2~E_{total}}{f\mathcal{V}} \label{eq: P_thermo}
\end{eqnarray}

From the above discussion, it can be deduced that for the time duration, when translational kinetic energy is less than its value at equilibrium, the mechanical pressure is also less than the mechanical pressure at equilibrium (i.e., the thermodynamic pressure). In case of compression, the situation would be reverse. Since, the energy supplied to the gas first goes to translational mode, instantaneous translational kinetic energy is more than its value at equilibrium. Therefore, the mechanical pressure is also higher than the mechanical pressure at equilibrium (i.e., the thermodynamic pressure).

\subsubsection{Intrinsic bulk viscosity}


This mechanism was first proposed by Hall \cite{hall1948origin} in 1948. Consider a compression of dense fluid such as liquid water. During compression, two different processes take place simultaneously. The first one is that molecules are brought uniformly closer together (just like zoom in/out of a computer graphics). It can be called molecular compression, and it is an almost instantaneous process. However, even if the molecular arrangement before the compression was a stable one, i.e., the one with minimum intermolecular potential energy, the molecular compression alone does not guarantee that the resulting molecular arrangement would also be a stable one. The reason for this is the nature of the intermolecular interactions, e.g., Lennard--Jones potential or hydrogen bonds.  Therefore, a second process also happens simultaneously. In this process, the molecules are rearranged or repacked more closely to achieve a more stable configuration. Hall identified this process as configurational or structural compression. This process involves the breaking of intermolecular bonds (e.g., hydrogen bonds) \cite{yahya2020molecular} or flow past energy barriers, which stabilizes the equilibrium configuration. This is a finite rate process. Thus it is of relaxational nature and is a source of nonequilibrium. This mechanism of bulk viscosity is present in all fluids including monatomic gases. Hence, monatomic gases at atmospheric conditions have a small ($\mathcal{O}(10^{-10})$~Pa~s) \cite{sharma2023bulkviscosityofmonatomicgases} but non-zero bulk viscosity. It should also be noted that at hypothetical dilute gas conditions, the bulk viscosity of monatomic gases is considered to be absolute zero.

Monatomic gases in their dilute gas limit do not possess any appreciable inter-molecular potential energy, therefore have negligible intrinsic bulk viscosity. Moreover, because of monatomic nature, they do not possess any internal degrees of freedom. Therefore, the mechanical pressure at any instant will always be equal to the thermodynamic pressure. Hence, we can expect them to exhibit negligible bulk viscosity. Both theories and experiments~\cite{prangsma1973ultrasonic,chapman1990mathematical} also confirm the same. 
%
%
However, it is possible that factors other than rotation/vibration and potential energy, like electronic excitation or chemical reaction~\cite{landau1959lifshitz} can also cause non-equilibrium in dilute monatomic gases. For instance, Istomin \emph{et al.} \cite{istomin2017transport} has shown that the bulk viscosity is not zero in electronically excited monatomic gases at temperatures higher than 2000 K. 

\section{Methods for determination of bulk viscosity} \label{Sec: Methods}

Unlike shear viscosity, determination of bulk viscosity has always remained a challenging task. We present a brief summary of various approaches available for estimation of bulk viscosity, $\mu_b$, of non-relativistic classical fluids, and  QGP and hadronic matter in Secs.~\ref{Sec: Classical-fluids} and \ref{Sec: QGP Methods}, respectively, with a particular focus on the former.

\subsection{Determination of bulk viscosity of classical fluids} \label{Sec: Classical-fluids}

A schematic overview of the methods available in the literature for the determination of bulk viscosity is shown in Fig. \ref{fig:surveyofexistingmethods}. A brief discussion on these methods is given below. 

\begin{figure}
	\centering
	\includegraphics[width=\linewidth, trim={2.5cm 18cm 3.6cm 2.5cm}]{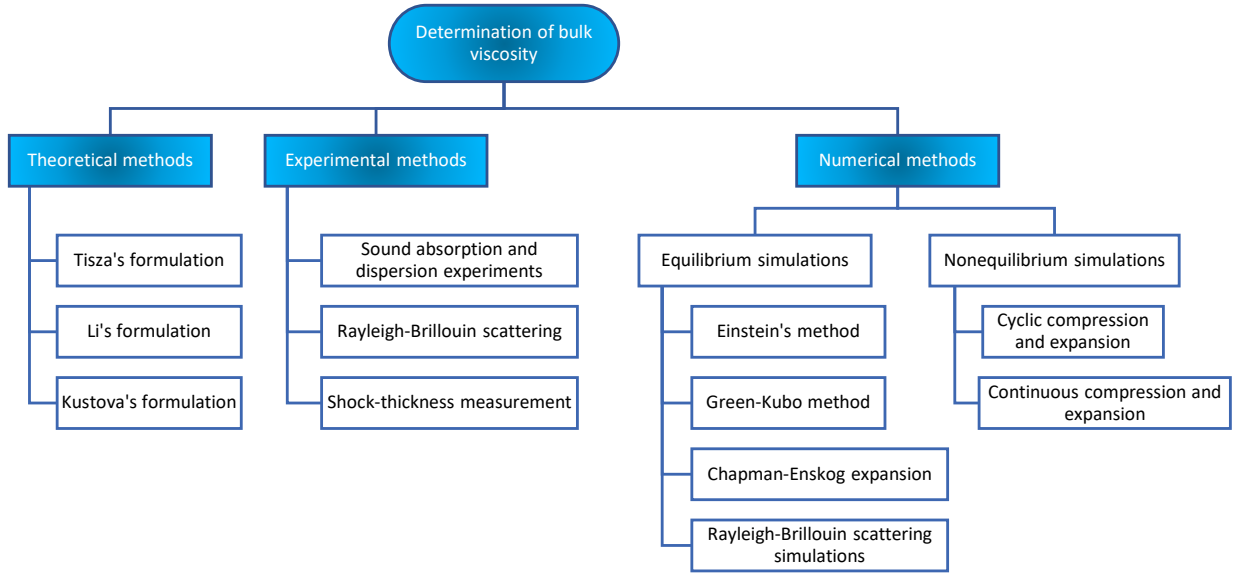}
	\caption{A survey of available methods for estimation of bulk viscosity.}
	\label{fig:surveyofexistingmethods}
\end{figure}

\subsubsection{Theoretical methods} \label{sec:Theoretical}

Theoretically, bulk viscosity of dilute gases can be related to the relaxation time of equilibration processes by Tisza's \cite{tisza1942supersonic} formulation, given as follows:
\begin{equation}
\label{eq:Tisza's_Formula}
\mu_b = \rho_{eq}\, a^2~ \frac{(\gamma -1)}{\gamma} ~\sum_{i}\frac{c_{v,i}}{c_v}\tau_i
\end{equation}
where $\rho_{eq}$ is density of gas at equilibrium, $a$ is speed of sound in absence of viscosity, $\gamma$ is the ratio of specific heats at equilibrium, $c_{v,i}$ is heat capacity of $i^{th}$ internal mode, $c_v$ is total heat capacity of the gas, $\tau_i$ is the relaxation time of that internal mode, and the summation is performed over all internal degrees of freedoms (i.e., rotational, vibrational). However, the applicability of this expression is limited to the low frequency regime where $\omega \tau << 1$; $\omega$ being the frequency of sound wave. \cite{meador1996bulkviscosity-fact-or-fiction}.
\newline
Li {\em et al.} \cite{li2017continuum} related bulk viscosity to bulk modulus and relaxation time as follows:
\begin{eqnarray}
\mu_b = K \tau_{tot}
\end{eqnarray}
where, $K$ is bulk modulus of the fluid, defined as $K=- \mathcal{V} ({\partial P}/{\partial \mathcal{V}})$, and $\tau_{tot}$ is total average relaxation time of internal energy in all excited modes.


Kustova \emph{et al.} \cite{kustova2019relaxation} questioned a priory split of bulk viscosity in rotational and vibrational components as done by Tisza\cite{tisza1942supersonic} and Cramer \cite{cramer2012numerical}.  They argued that this splitting lead to overprediction of bulk viscosity at low temperatures and bulk to shear viscosity ratio of $\sim2000$ for carbon dioxide at low temperatures is unjustified. They first derived a expression for bulk viscosity using modified Chapman-Enskog method \cite{nagnibeda2009non} under the assumption of local thermodynamic equilibrium
\begin{equation}
	\mu_b = \displaystyle \frac{R \, c_{int}}{c_v^2} \, p \, \tau_{int} = \frac{(\gamma-1)^2}{R} \, c_{int}\,  p \, \tau_{int}
	\label{eq:Kustova-mub-1}
\end{equation}
where $c_{int}$ and $\tau_{int}$ are total heat capacity of internal degrees of freedoms and corresponding relaxation time, respectively. Then following the work by Mason and Monchick\cite{mason1962heat}, they assumed that collisions with simultaneous exchange of rotational and vibrational energies are rare. 
By invoking this assumption, it can be shown that
\begin{equation}
	\frac{c_{int}}{\tau_{int}} = \frac{c_{rot}}{\tau_{rot}} + \frac{c_{vib}}{\tau_{vib}}
	\label{eq:Kustova-mub-2}
\end{equation}
By using Eq~\eqref{eq:Kustova-mub-1} and \eqref{eq:Kustova-mub-2}, bulk viscosity can be expressed as 
\begin{equation}
	\mu_b = pR\left(\frac{c_{int}}{c_v}\right)^2\left(\frac{c_{rot}}{\tau_{rot}} + \frac{c_{vib}}{\tau_{vib}}\right)^{-1}
	\label{eq:Kustova-mub-final}
\end{equation}
On the basis of above formulation, Kustova \emph{et al.} also deduced that bulk to shear viscosity ratio for CO$_2$ at 300-1000~K should be in the range 1 to 3.

\subsubsection{Experimental methods}

The experimental determination of bulk viscosity is not as straightforward as shear viscosity and usually based on indirect techniques, such as absorption and dispersion of sound wave, and Rayleigh-Brillouin scattering \cite{hoover1980bulk,zaheri2007theoretical,graves1999bulk}.

Absorption of the sound wave is the additional decrease in intensity with distance, over and above the geometric reduction caused by the inverse square law. It has been found that the experimentally observed absorption is much higher than the predictions based on the theory that accounts only for classical absorption, i.e., absorption due to shear viscosity, thermal conductivity, and thermal radiation. Since this excess absorption cannot be attributed to dissipation phenomenon because of translational motion of molecules (i.e., shear viscosity, heat conduction), it is assumed that this excess absorption is because of bulk viscosity \cite{karim1952second}. The absorption of sound is characterized by absorption coefficient ($\alpha$), and it is related to bulk viscosity, $\mu_b$, as following \cite{prangsma1973ultrasonic}:
\begin{equation}
\label{eq:alpha}
\frac{\alpha P_{eq}\,}{\omega^2}=\frac{2 \pi^2}{\gamma a} \displaystyle \left[ \frac{4}{3}\mu + \frac{(\gamma - 1)^2}{\gamma}\frac{M \kappa}{R}+\mu_b \right]
\end{equation}
where, $P_{eq}$ is equilibrium pressure, 
$M$ is molar mass, $\kappa$ is thermal conductivity, and $R$ is gas constant. 

However, the assumption that the excess absorption is due to bulk viscosity can only be examined when some direct measurements of bulk viscosity from an independent method are made and values are compared \cite{truesdell1954present, karim1952second}. Furthermore, this approach of measuring bulk viscosity is susceptible to considerable errors since it involves subtraction of classical absorption coefficient from total absorption coefficient to get absorption due to bulk viscosity. For the calculation of classical absorption, the use of $\mu$ and $\kappa$ taken from different sources also introduces error in estimates of bulk viscosity made using this method \cite{graves1999bulk}.
 
Alternatively, bulk viscosity can also be measured by sound dispersion experiments. Dispersion of sound causes speed of sound to be frequency dependent, and this dependency is given as follows \cite{winter1967high-temperature-ultasonic-measurements}- 
\begin{equation}
a^2 = a_{0}^{2}+(a_{\infty}^{2}-a_{0}^{2}) \frac{\omega^2 \tau^2}{1+\omega^2 \tau^2}
\end{equation} 
where $a$ is the speed of sound at frequency $\omega$, $a_0$ and $a_\infty$ are speed of sound for very low and very high frequencies, respectively. The obtained relaxation time then can be related to bulk viscosity.

Pan \emph{et al.} \cite{pan2005power} suggested that bulk viscosity of dilute gases can also be measured using Coherent Rayleigh -- Brillouin Scattering (CRBS). In this technique, gas density perturbations are generated and measured using laser beams. The experimentally observed scattering profile is then compared with that obtained from the theoretical models to get transport coefficients, including bulk viscosity. However, in contrast to acoustic experiments, which measure bulk viscosity at megahertz frequencies, these experiments measure bulk viscosity in gigahertz frequencies. Because of this reason, a significant difference between the estimated values from the two above mentioned methods is usually observed \cite{pan2005power, pan2004coherent, vieitez2010coherent,gu2013temperature,gu2014systematic,meijer2010coherent, dukhin2009bulk}.

Furthermore, Emanuel \emph{et al.}~\cite{emanuel1994linear} has deduced that for dense polyatomic gases, the density-based thickness of shock wave consists of many thousands of mean-free paths, and varies linearly with the ratio $\mu_b/\mu$. Thus, ideally, the experimentally measured shock wave thickness can be used to calculate bulk viscosity; nevertheless, we could not find any experimental implementation of this technique in the literature. 

\subsubsection{Computational methods}

The methods employed for the determination of transport properties using numerical simulations are broadly classified in two categories-- non-equilibrium simulation based methods (e.g., non-equilibrium molecular dynamics (NEMD)), and equilibrium simulation based methods (e.g., equilibrium molecular dynamics (EMD)) (see Fig. \ref{fig:surveyofexistingmethods}). In the former approach, the non-equilibrium responsible for the desired transport property is produced at the microscopic level, and then the transport property is related to other variables as in physical experiments. The latter approach uses relations, such as Green-Kubo relation\cite{green1954markoff,kubo1957statistical}, Einstein relation \cite{helfand1960transport, viscardy2007transport}, or expressions derived from the Chapman-Enskog expansion\cite{bruno2019direct, bruno2015oxygen, taxman1958classical, chapman1990mathematical,wang1951transport}.


\paragraph*{Equilibrium based methods}

The Green-Kubo method uses the Green-Kubo relations for calculation of transport properties \cite{green1954markoff,kubo1957statistical}. For shear viscosity, $\mu$, and bulk viscosity, $\mu_b$, these relations are given as follows:
\begin{equation}
\label{eq:Green-Kubo_shear-viscosity}
\mu = \frac{\mathcal{V}}{k_B T} \int_{0}^{\infty} \langle P_{i j}(t)~P_{i j}(0) \rangle dt
\end{equation}
where $k_B$ is the Boltzmann constant, $T$ is temperature, $P_{i j}(t)$ denotes the instantaneous value of $ij^{\text{th}}$ off-diagonal element of the pressure tensor at a time $t$, and the angle bracket indicates the ensemble average. Further, to reduce the statistical error in calculation of $\mu$, averaging is performed over three different values obtained from three different components of pressure tensor viz., $P_{ij}$, $P_{jk}$, and $P_{ki}$.
\begin{equation}
\mu_b = \frac{\mathcal{V}}{k_B T} \int_{0}^{\infty} \langle \delta P(t)~\delta P(0) \rangle dt
\end{equation}
Here, $P(t)$ is instantaneous value of the average of three diagonal terms of pressure tensor at a time $t$, i.e., $P(t)=\frac{1}{3}[P_{ii}(t)+P_{jj}(t)+ P_{kk}(t)]$. The fluctuations, $\delta P(t)$, is aberration of mean pressure from equilibrium pressure, i.e., $\delta P(t)=P(t)-P_{eq}$; where $P_{eq}$ is equilibrium pressure of the system, and it is calculated by time average of P(t) over a long time. 

The Green-Kubo method is a robust way to measure transport coefficients, but sometimes, they suffer from several issues. For example, for the correct estimation of viscosity coefficients, the auto-correlation function should decay to zero with time. In such a case, the integral of auto-correlation function would reach a constant value. Nevertheless, it does not necessarily happen in practice. The auto-correlation function might show either long-time tails or fluctuations \cite{zhang2015reliable-time-decomposition-method}. Further, viscosities should be estimated from the region of the graph of the integral (of auto-correlation function) vs. time, when it reaches a constant value. First of all, it is difficult to identify such a region, and second, even if we can identify such a region, there will always be arbitrariness in the value of viscosity because of ambiguity in determining the cut-off time \cite{zhang2015reliable-time-decomposition-method}.

The Einstein relations \cite{meier2004transport-1viscosity, meier2005transport} also find their origin in linear response theory. These relations relate shear and bulk viscosity to the slope of generalized mean-squared displacement functions as given below 
\begin{align}
	\mu &= \frac{\mathcal{V}}{2k_BT} \lim_{t \to \infty }\frac{d}{dt} \left\langle \left[  \frac{m}{\mathcal{V}}\sum_{n=1}^{N} [v_{n,i}(t) r_{n,j}(t) - v_{n,i}(0) r_{n,j}(0)        ] \right]^2 \right\rangle \\
	\mu_b &= \frac{\mathcal{V}}{k_BT} \lim_{t \to \infty }\frac{d}{dt} \left \langle \left[  \frac{m}{3\mathcal{V}}\sum_{n=1}^{N} [\vec{v}_{n}(t) \cdot \vec{r}_{n}(t) - \vec{v}_{n}(0)\cdot  \vec{r}_{n}(0)] - P_{eq}t \right]^2 \right \rangle .
\end{align}
Here, $N$ is total number of particles, $m$ is particle mass, $\vec{r}_n (t)$ and $\vec{v}_n (t)$ are the position and velocity vectors of $n^{th}$ particle at time $t$, and $\vec{r}_{n,i} (t)$ and $\vec{v}_{n,j} (t)$ are the Cartesian components of position and velocity vectors in direction $i$ and $j$. 

The Einstein relations are theoretically equivalent to Green-Kubo relations. However, they can not be directly implemented in molecular dynamics (MD) simulations with periodic boundaries\cite{erpenbeck1995einstein}. The Einstein relations implicitly assumes the particles to follow a continuous trajectory. Whereas, in case of periodic boundaries, particles regularly exit from one face and enter from opposite face of the domain, therefore, the trajectory becomes discontinuous. This problem is overcome by using modified Einstein relations where the generalized displacement functions are replaced with time integral of pressure tensor components. The modified Einstein relations are given as follow
\begin{align}
	\mu = \frac{\mathcal{V}}{2k_bT}\lim_{t \to \infty } \frac{d}{dt} \left\langle \left[ \int_{0}^{t} P_{ij}(t)dt \right]^2\right \rangle \\
	\mu_b = \frac{\mathcal{V}}{2k_bT}\lim_{t \to \infty } \frac{d}{dt} \left\langle \left[ \int_{0}^{t} \delta P(t)dt \right]^2\right\rangle .
\end{align}

The Chapman-Enskog theory based expressions for transport coefficients of dilute polyatomic gases, derived by Taxman \cite{taxman1958classical}, requires the calculation of collision integrals. These collision integrals are usually evaluated through the Monte Carlo quadrature method\cite{hellmann2008calculation}, in which proper equilibrium distribution function is used to sample pre-collision state of molecules, and then post-collision states are calculated by solving molecular trajectory classically\cite{bruno2015oxygen}.
The Chapman-Enskog method is computationally more efficient than the Green-Kubo method, however, the applicability of this method depends upon the availability of explicit expressions, whereas the Green-Kubo method, though less efficient, does not have any such constraint \cite{bruno2015oxygen}.

In another approach, transport coefficients can also be estimated by numerical simulation of the Rayleigh-Brillouin scattering experiment and its coherent version, coherent Rayleigh Brillouin scattering. In this method, a gas is simulated at equilibrium, and density fluctuations are sampled as time-series data. The discrete power spectrum is then obtained using the square of the Fourier transformation of this time series \cite{bruno2019direct, baras1995particle, cornella2012experimental}. 

All the three above-mentioned methods, i.e., Chapman-Enskog expressions, Green-Kubo relations, and simulation of Rayleigh-Brillouin scattering can be employed in either MD and classical trajectory direct simulation Monte Carlo (CT-DSMC) simulation framework. More details on these three methods are available in Ref.\cite{sharma2019estimation, bruno2019direct, dukhin2009bulk, bruno2015oxygen}.

\paragraph*{Non-equilibrium based methods}

Non-equilibrium based methods measures transport coefficients by directly measuring the gradient of the corresponding parameter. Such methods are well developed for shear viscosity, heat conductivity, and mass diffusivity. However, for bulk viscosity, historically it has been seen that implementation of such a method is a challenging task. To the best of our knowledge, only one attempt has been made so far to use non-equilibrium based methods for the estimation of bulk viscosity. In this work, Hoover {\em et al.} \cite{hoover1980bulk} cyclically compressed and expanded the fluid in molecular dynamics framework in the following manner to produce measurable effects:
\begin{equation}
	\label{eq:hoover_deformation}
	L/L_0=1+\xi \sin(\omega t)
\end{equation}
where, $L_0$ is mean length, $L$ is instantaneous length of the cubic simulation domain, $\xi$ is strain amplitude, and $\omega$ is the frequency describing the linearized strain rate.
\begin{equation}
	\dot{\epsilon}=\xi \omega ~\cos(\omega t)
\end{equation}
As the linearized strain rate ($\dot{\epsilon}$) approaches zero, authors expected the average pressure of the system to deviate from equilibrium pressure by $-3\xi \omega \mu_b \cos(\omega t)$. Also, if the deformation given by Eq.~(\ref{eq:hoover_deformation}) takes place through the mechanism of external work, then the lost work due to irreversible heating will give rise to energy increase per cycle, $\Delta E_{\text{per cycle}}$, (the cycle time being $2\pi/\omega$), which is equal to 
\begin{equation}
	\Delta E_{\text{per cycle}} = (2\pi/\omega) 9 \xi^2 \omega ^2 \mu_b \mathcal{V}/2
	\label{eq:energy_absorbed_in_cyclic_comp_and_exp}
\end{equation}

Based on the measurement of average pressure and rise in the energy of the system, authors estimated bulk viscosity of a soft sphere fluid modeled with potential $\phi (r) = \epsilon (\sigma/r)^{12}$.

All of the methods discussed above have one or more limitations. The sound absorption and dispersion method give frequency-dependent bulk viscosity values. However, since bulk viscosity is a transport coefficient, it should depend only on the state of fluid, not on the process it is undergoing. The optical method, i.e., Rayleigh-Brillouin scattering, can not account for vibrational effects, since the frequency used is of GHz range, and 1/frequency becomes smaller than vibrational relaxation time. In the second vertical, i.e., numerical methods, the Green-Kubo method sometimes faces difficulties in convergence due to long-time tails. Chapman-Enskog method is very robust and fastest among all numerical methods. However, Chapman-Enskog relations are difficult to obtain except for simple cases like monatomic and diatomic gases. In the second sub-vertical, there is only one nonequilibrium method, i.e., cyclic compression and expansion by Hoover {\emph{et al.}}\cite{hoover1980bulk}. There is only one application available in the literature\cite{hoheisel1987bulk}. Since this method uses energy absorbed over many compression and expansion cycles to estimate bulk viscosity, it may also give frequency-dependent bulk viscosity values. Moreover, this method does not give much insights associated with physics of one single cycle. Therefore, to address these issue, Sharma {\emph{et al.}} \cite{sharma2019estimation} proposed a first principle based continuous compression/expansion method which can determine bulk viscosity directly from the difference between mechanical ($P_{mech}$) and thermodynamic pressure ($P_{thermo}$). 

Their method uses numerical measurement of $P_{mech}$ and $P_{thermo}$ in an NEMD simulation of an expanding fluid, and then relates the bulk viscosity to them by the relation $\mu_b =(P_{thermo}-P_{mech})/{\nabla\cdot\vec{u}}$, where ${\nabla\cdot\vec{u}}$ is the controlled rate of expansion of the fluid per unit volume. The key success of this method was that it, being inherently based on expansion/compression of the gas, allowed them to investigate the effects of nonequilibrium on bulk viscosity, e.g., variation of bulk viscosity with magnitude ($|\nabla \cdot \vec{u}|$) and direction of volumetric change (i.e., expansion vs. compression). Hence, this method enabled them to gain detailed insights of associated flow physics. 

 
\subsection{Determination of bulk biscosity of QGP and hadronic matter} \label{Sec: QGP Methods}

For the estimation of transport coefficient of QGP and hadronic matter, two standard approaches are mainly used -- the Boltzmann equation based relaxation time approximation (RTA) approach and the linear response theory based Kubo/Green-Kubo formulation. A brief review of these methods can be found in Ref. \cite{demir2010extraction, oertzen1990transport, saha2017comparative}. A vast amount of research has been done on this topic. Here we summarize only some of the key contributions in this field. Gavin \cite{gavin1985transport} used the well known non-relativistic form of the Boltzmann equation to calculate the transport coefficients for both the QGP and hadronic matter using the relaxation time approximation (RTA) method. Prakash \emph{et al.} \cite{prakash1993non,prakash1993fast} studied the equilibration of hot hadronic matter in the framework of relativistic kinetic theory. They calculated transport coefficients considering only elastic collisions in the dilute gas limit using extended Chapman-Enskog formalism. For a general review of the relativistic kinetic theory, reader may refer to the classical text by Groot \emph{et al.} \cite{groot1980relativistic}.
Chakraborty \emph{et al.} \cite{chakraborty2011quasiparticle} extended the classical works of Prakash \emph{et al.} \cite{prakash1993non,prakash1993fast} and Gavin\cite{gavin1985transport}, and presented a theoretical framework for the calculation of shear and bulk viscosity of hot hadronic matter. Their work accounted for not only inelastic collisions but also formation and decay of resonances, temperature-dependent mean fields, and temperature-dependent masses.
Demir \emph{et al.} \cite{demir2009extracting, demir2009shear,demir2010extraction} carried out Ultra-relativistic Quantum Molecular Dynamics (UrQMD) simulations of hadronic media, and calculated bulk viscosity using both Green-Kubo method and relaxation time approximation.  

\section{Applications of bulk viscosity}
\label{Sec:Applications}
To evaluate the validity of Stokes' hypothesis, let us consider a supersonic flow of air and analyze the terms of the equation $ P_{mech} = P_{thermo} - \mu_b \nabla\cdot\vec{u}$. For bulk viscosity effects to be significant, $\mu_b \nabla\cdot\vec{u}$ should be comparable to $P_{thermo}$. Assuming pressure is $\sim 10^5$~Pa and $\mu_b$ is $\sim 10^{-5}$~Pa~s, even if $\nabla\cdot\vec{u}$ is as high as $ \approx 10^4 $~s$^{-1}$, the difference between mechanical and thermodynamic pressure would be just 0.1~Pa. This difference $ P_{thermo}-P_{mech}$ = 0.1 Pa is six orders of magnitude smaller than the $P_{thermo}$. Therefore, in most of the commonly encountered flow problems, it is safe to assume bulk viscosity to be zero. However, bulk viscosity effects may become important when either the $\nabla\cdot\vec{u}$ is very high (e.g., inside a shock wave), or when fluid is compressed and expanded in repeated cycles such that the cumulative effect of the small contributions from each cycle is no more negligible (e.g., sound wave) \cite{buresti2015note}, or when the atmosphere consists of the majority of those gases, such as CO$_2$, which exhibit a large bulk viscosity \cite{emanuel1992effect}, or when results of interest might get affected by even small disturbances, e.g., the study of Rayleigh-Taylor instability \cite{sengupta2016roles}. In such cases, it becomes necessary to account for the bulk viscosity terms in the Navier--Stokes equation. 

Several researchers have investigated the effects of the incorporation of bulk viscosity in analytical or CFD studies of various flow scenarios. Emanuel \emph{et al.} \cite{emanuel1990bulk, emanuel1992effect, emanuel1994linear, emanuel1998bulk} reviewed bulk viscosity and suggested that the effects of bulk viscosity should be accounted for in the study of high-speed entry into planetary atmospheres. They observed that the inclusion of bulk viscosity could significantly increase heat transfer in the hypersonic boundary layer\cite{emanuel1992effect}. Chikitkin \cite{chikitkin2015effect} studied the effects of bulk viscosity in flow past a spacecraft. They reported that the consideration of bulk viscosity improved the agreement of velocity profile and shock wave thickness with experiments. Shevlev \cite{shevelev2015bulk} studied the effects of bulk viscosity on CO$_2$ hypersonic flow around blunt bodies. The conclusions of their study were in line with that of Emanuel. They suggested that incorporation of bulk viscosity may improve predictions of surface heat transfer and other flow properties in shock layer.

Elizarova \emph{et al.} \cite{elizarova2007numerical} and Claycomb \emph{et al.} \cite{claycomb2008extending} carried out CFD simulations of normal shock. They found that including bulk viscosity improved the agreement with experimental observations for shock wave thickness. A recent study by Kosuge and Aoki \cite{kosuge2018shock} on shock-wave structure for polyatomic gases also confirms the same. Bahmani \emph{et al.} \cite{bahmani2014suppression} studied the effects of large bulk to shear viscosity ratio on shock boundary layer interaction. They found that a sufficiently high bulk to shear viscosity ratio can suppress the shock-induced flow separation. Singh and Myong \cite{singh2017computational} studied the effects of bulk viscosity in shock-vortex interaction in monatomic and diatomic gases. They reported a substantially strengthened enstrophy evolution in the case of diatomic gas flow. Singh \emph{et al.} \cite{singh2021impact} investigated the impact of bulk viscosity on the flow morphology of a shock-accelerated cylindrical light bubble in diatomic and polyatomic gases. They found that the diatomic and polyatomic gases have significantly different flow morphology than monatomic gases. They produce larger rolled-up vortex chains, various inward jet formations, and large mixing zones with strong, large-scale expansion. Touber \cite{touber2019small} studied the effects of bulk viscosity in the dissipation of energy in turbulent flows. He found that large bulk-to-shear viscosity ratios may enhance transfers to small-scale solenoidal kinetic energy, and therefore, faster dissipation rates. Riabov \cite{riabov2019limitations} questioned the ability of bulk viscosity to model spherically-expanding nitrogen flows in temperature range 10 to 1000~K by comparing results to Navier--Stokes equations to relaxation equation. He reported that the bulk viscosity approach predicts much thinner spherical shock wave areas than those predicted by relaxation equations. Moreover, the distributions of rotational temperature along the radial direction predicted by the bulk viscosity approach had neither any physical meaning nor matches with any known experimental data for expanding nitrogen flows. 

Fru \emph{et al.} \cite{fru2011direct} performed direct numerical simulations (DNS) study of high turbulence combustion of premixed methane gas. They found that the incorporation of bulk viscosity does not impact flame structures in both laminar and turbulent flow regimes. Later, the same group extended their study to other fuels, viz., hydrogen and synthetic gas. In this study \cite{fru2012impact}, they found that though flame structures of methane remained unchanged before and after incorporation of bulk viscosity, the same for hydrogen and syngas showed noticeable modifications. 
Sengupta \emph{et al.} \cite{sengupta2016roles} studied the role of bulk viscosity on Rayleigh Taylor instability. They found that the growth of the mixing layer depends upon bulk viscosity. Pan \emph{et al.} \cite{pan2017role} has shown that bulk viscosity effects cannot be neglected for turbulent flows of fluids with high bulk to shear viscosity ratio. They found that bulk viscosity increases the decay rate of turbulent kinetic energy. Boukharfane \emph{et al.} \cite{boukharfane2019role} studied the mechanism through which bulk viscosity affects the turbulent flow. They found that the local and instantaneous structure of the mixing layer may vary significantly if bulk viscosity effects are taken into account. They identified that the mean statistical quantities, e.g., the vorticity thickness growth rate, do not get affected by bulk viscosity. On this basis of their study, they concluded that results of refined large-eddy simulations (LES) might show dependence on the presence/absence of bulk viscosity, but Reynolds-averaged Navier--Stokes (RANS) simulations might not as they are based on statistical averages.

Connor \cite{szemberg2018bulk} studied the effects of bulk viscosity in the compressible turbulent one-, two-, and three-dimensional Couette flows through DNS simulations. The objective of the study was to test whether invoking the Stokes' hypothesis introduces significant errors in the analysis of compressible flow of solar thermal power plants,  and carbon capture and storage (CCS) compressors. They found that most of the energy is contained in the solenoidal velocity for both CCS and concentrated solar power plants. Therefore, assuming bulk viscosity to be zero does not produce any significant errors, despite the compressors operating at supersonic conditions. However, bulk viscosity effects may become significantly close to the thermodynamic critical point.

Billet \emph{et al.} \cite{billet2008impact} showed that the inclusion of bulk viscosity in CFD simulations of supersonic combustion modifies the vorticity of the flow. Lin \emph{et al.} \cite{lin2017high, lin2017bulk} have shown that acoustic wave attenuation in CFD simulations can be accounted for by incorporating bulk viscosity. Nazari \cite{nazari2018important} studied the influence of liquid bulk viscosity on the dynamics of a single cavitation bubble. They reported that bulk viscosity significantly affects the collapse phase of the bubble at high ultrasonic amplitudes and high viscosities. High bulk viscosity values also altered the maximum pressure value inside the bubble. 

The classical Navier--Stokes equation loses its accuracy as the deviation from equilibrium increases, or the extent of nonequilibrium is high. This usually happens in the study of high-temperature gas dynamics and rarefied gas dynamics. There are primarily three models used in the continuum framework to model nonequilibrium, viz., state-to-state, multi-temperature, and one-temperature model. The state-to-state model is computationally most costly and can be used in systems that are far from equilibrium and have coupled fluid, thermal and chemical kinetics. In comparison, the one-temperature model is computationally most simple and suitable for near-equilibrium systems. In one temperature model, both rotational and vibrational relaxations are accounted for using bulk viscosity coefficient. On the other hand, in the state-to-state approach, the vibrational chemical kinetics is described by master equations for populations of vibrational states, and the fast rotational relaxation is accounted for through bulk viscosity parameters. Similarly, in the multi-temperature model, vibrational kinetics is governed by relaxation equations for various vibrational modes, and the rotational relaxation is modeled through bulk viscosity. For a more detailed discussion on these models, the reader is referred to the Ref.~\cite{kustova2019relaxation}.

In addition to these classical fluid-dynamics scenarios, bulk viscosity may also play an important role in several cosmological phenomena, e.g., damping of vibrations created during the formation of a new neutron star, and growth of gravitational--wave instability in rapidly rotating neutron star~\cite{sawyer1989bulk, dong2007bulk}. Origin of bulk viscosity in these circumstances is primarily due to the chemical nonequilibrium caused by nuclear reactions. Bulk viscosity along with other transport properties is also of central importance to the space-time description of the heavy-ion collision experiments being conducted at the Brookhaven National Laboratory's Relativistic Heavy Ion Collider (RHIC) and CERN's Large Hadron Collider (LHC). One of the primary objectives of these experiments was the formation and investigation of Quark-Gluon-Plasma (QGP), the state of the matter of the Early Universe (first 30 $\mu$s after the Big Bang). \cite{demir2010extraction}
%

\section{Concluding remarks and future scope}
Although of great importance in several fluid mechanical phenomena, bulk viscosity is still one of the gray areas of fluid mechanics. Despite the fact that numerous theoretical, experimental, and computational studies have been carried out in the past, there is still lack of well-established values for bulk viscosity of common gases such as nitrogen, oxygen, and carbon dioxide. For most gas mixtures, data is not available at all, or if available, experimental or numerical, is spread in a wide range. Therefore, studies that determine the bulk viscosity of common fluids are needed to achieve a common consensus on accepted values.

\section*{Disclaimer}
The contents of this article is primarily based on Introduction chapter of the thesis: \fullcite{sharma2022nature}\\

\appendix
\section*{Appendix A: Selected references for further reading}
Reader is redirected to following resources for further understanding of concepts and historic developments in the subject of bulk viscosity. Enumerated items are arranged in the suggested order of reading.
\begin{itemize}
	\item For first introduction to the subject:
	\begin{enumerate}
		\item \fullcite{gad1995questions}
		\item \fullcite{buresti2015note}
	\end{enumerate}
	
	\item For a detailed understanding:
	\begin{enumerate}
		\item \fullcite{sharma2022nature}
		\item \fullcite{landau1959lifshitz} (Chapter~2, Sec~15 and Chapter-8, Sec~81)
		\item \fullcite{herzfeld2013absorption}
	\end{enumerate}
	
	\item For historic developments:
	\begin{enumerate}
		\item \fullcite{hall1948origin}
		\item \fullcite{karim1952second}
		\item \fullcite{rosenhead1954discussion}
		\item \fullcite{davies1954kinetic}
		\item \fullcite{borovinskii1988concept}
		\item \fullcite{truesdell1954present}
		\item \fullcite{emanuel1990bulk}
		\item \fullcite{meador1996bulkviscosity-fact-or-fiction}
		\item \fullcite{dong2007bulk}
	\end{enumerate}
	
	\item For complete treatment of equation of motion of viscous fluids without Stokes' hypothesis:
	\begin{enumerate}
		\item \fullcite{emanuel2000analytical}
	\end{enumerate}
	
	\item For fundamentals of equilibrium statistical mechanics:
	\begin{enumerate}
		\item \fullcite{bagchi2018statistical}
		\item \fullcite{allen2017computer}
	\end{enumerate}
\end{itemize}

\printbibliography
\end{document}